\documentclass{article}
\usepackage{graphicx} 
\usepackage{amsmath}
\usepackage{appendix}
\usepackage{array}
\usepackage{longtable}
\usepackage[authoryear,round]{natbib}
\usepackage[hidelinks]{hyperref}
\usepackage{booktabs}
\usepackage[table]{xcolor}
\usepackage[acronym]{glossaries}
\makenoidxglossaries
\usepackage[margin=1in]{geometry}
\usepackage{microtype}
\usepackage{placeins}

\usepackage{xurl}

\newacronym{oos}{OOS}{out-of-sample}
\newacronym{vwa}{VWA}{volume-weighted average}
\newacronym{vwga}{VWGA}{volume-weighted geometric average}
\newacronym{rmspe}{RMSPE}{root mean squared percentage error}
\newacronym{cdr}{CDR}{claims development result}
\newacronym{maave}{MAAvE}{mean absolute actual versus expected}
\newacronym{mapave}{MAPAvE}{mean absolute percentage actual versus expected}
\newacronym{mpe}{MPE}{mean percentage error}

\title{A Multiplicative Loss Function for Chain Ladder}

\author{
  James Grove \\
  \small Dynamo Analytics \\
  \small \texttt{james.grove@dyna-mo.com}
  \and
  Stephan Marais\thanks{Corresponding author.} \\
  \small Dynamo Analytics \\
  \small \texttt{stephan.marais@dyna-mo.com}
}
\date{August 2026}

\begin{document}

\maketitle

\begin{abstract}
Reserving models increasingly rely on loss-based estimation, where the loss function encodes the assumed error structure. Mack demonstrated this for the chain ladder, showing that the volume-weighted average estimator minimises a volume-weighted squared-error loss function that is additive in successive claim developments. This paper instead considers a multiplicative error structure and proposes the corresponding volume-weighted squared log-error loss function. Adapting Mack's distribution-free framework, we show that this loss function is minimised by the volume-weighted geometric average of the individual development ratios. This provides practitioners with an alternative estimator of development ratios for chain-ladder-based models, and a candidate loss function for machine-learning-based reserving models. We further show that the same estimator arises from two independent arguments: a stability requirement on successive ultimate loss projections, and maximum-likelihood estimation under a log-normal model. The estimator thus admits three complementary justifications: loss minimisation, reserve stability, and parametric likelihood. An out-of-sample study of 362 Schedule P company-line datasets supports the use of the proposed estimator in place of the volume-weighted average for the chain ladder, by showing that it improves predictive accuracy and reduces a slight over-prediction bias.
\end{abstract}

\vspace{1em}

\section{Introduction}

The chain ladder method predicts outstanding claims through successive multiplicative development ratios. However, the standard \gls{vwa} approach to estimating development ratios implicitly assumes additive errors in those successive developments \citep{ref:Mack1993,ref:mack1994}. This paper argues that a multiplicative error - an assumption with a long history in log-normal models \citep{ref:kremer1982,ref:hertig1985,ref:verrall1991,ref:geometricCL} - is a natural alternative consistent with the chain ladder's multiplicative structure, and explores its consequences within the distribution-free framework of \cite{ref:Mack1993}.

We ask whether a multiplicative error structure has merit and, if so, under what conditions, considering both theory and empirical performance.

Owing to its simplicity, the chain ladder is often viewed as a purely arithmetic method that extrapolates cohort-based development patterns from historical data. Expressing the chain ladder as a stochastic model makes its underlying assumptions explicit and provides a principled basis for parameter estimation. \cite{ref:mack1994} showed that choosing development ratios to best predict successive cash flows in the historical triangle yields the \gls{vwa} ratios, and that these assume a model error additive over successive developments. This error's variance depends on the origin period for a given development period.

This paper investigates fitting reserving models with a loss function derived from a multiplicative error assumption for successive claim developments, which amounts to working with the logarithms of the individual development ratios rather than the ratios themselves. Applying this within the chain ladder method, we extend the work of \cite{ref:Mack1993,ref:mack1994} to see whether estimating parameters under a multiplicative error produces accurate estimates and yields new insights. Because the chain ladder is so widely used in practice, any resulting insights are readily interpretable and implementable.

We distinguish key terminology here. The multiplicative structure of the chain ladder model, the assumed form of the error, and the loss function used for parameter estimation are closely related, but conceptually distinct.

Modelling claims development on the logarithmic scale is itself not new. \cite{ref:kremer1982} formulated a log-additive two-way model for incremental claims, \cite{ref:verrall1991} developed estimation for the log-normal chain ladder, and \cite{ref:hertig1985} modelled the logarithms of the individual development ratios as normal random variables. More recently, \cite{ref:geometricCL} derived a geometric chain ladder by assuming the incremental claims are log-normal. The model-based treatments in this literature are, however, fully parametric, whereas the present paper remains within the distribution-free framework of \cite{ref:Mack1993} and assumes only the first two moments of the error. For additional motivation and illustration of the underlying assumptions, we show how a parametric assumption can also lead to the same solution.

This paper makes the following contributions. We derive the \gls{vwga} estimator as the minimiser of a volume-weighted squared log-error loss function inspired by Mack's distribution-free framework \citep{ref:Mack1993}. We show that the same estimator arises independently from the requirement of stability in successive ultimate loss projections, and that it also follows from a parametric assumption about the individual development ratios' distribution. Finally, we provide empirical evidence, based on 362 Schedule P company-line datasets, each with a paid and an incurred triangle, demonstrating improved \gls{oos} predictive performance relative to the \gls{vwa} approach.

Loss reserving modelling techniques have advanced significantly since \cite{ref:Mack1993} showed how the chain ladder method can be expressed as a distribution free stochastic model. Many of these newer methods are inspired by statistical learning theory and require specifying a loss function to both evaluate predictive accuracy and train models. Common examples include the root mean squared error, the volume-weighted mean squared error, and the \gls{cdr} score \citep{ref:merz2008CDR,ref:CDRScore}. \cite{ref:mack1994} showed that the \gls{vwa} pattern minimises the volume-weighted mean squared error for the chain ladder method, providing an early example of loss-based estimation in reserving as seen in statistical learning theory. If the error structure of the chain ladder method can be improved, this may also inform the loss functions and underlying assumptions employed in these modern techniques. This highlights the relevance of the present work.

The remainder of the paper is structured as follows. We begin by reviewing the work of \cite{ref:Mack1993,ref:mack1994}, focusing on the derivation of the \gls{vwa} development ratios as parameter estimates for chain ladder. We then derive new parameter estimates under an explicit multiplicative error assumption, which yields the \gls{vwga} chain ladder method. To assess the validity and efficiency of using a multiplicative error versus an additive error, we compare the two methods in terms of their ability to predict \gls{oos} outcomes using the Schedule P public data from accident years 1998--2007 \citep{ref:ScheduleP}.

\section{Background - the additive-error baseline}\label{sec:background}

In this paper, we will use the following notation and conventions.
Cumulative total claims for each origin period $i$ and development period $j$ are denoted $C_{i,j},$ $i\in\{1,2,\ldots,I\},$ $j\in\{1,2,\ldots,J\}$ so that $I$ denotes the number of origin periods and $J$ the number of development periods of the triangle. Similarly, we have incremental cash flows for the triangle as $X_{i,j},$ $i\in\{1,2,\ldots,I\},$ $j\in\{1,2,\ldots,J\}$.

To simplify notation in this paper we assume $I=J$ so that the number of origins and developments are equal. We can then define \[\Delta^k=\{C_{i,j}: i + j \le k + 1\},\] as the set of all the cumulative cash flows at calendar period $k\in\{1,2,\ldots,J\}$, where $k$ indexes the calendar-period diagonal of the triangle. Similarly, let $\Delta^{-k}$ correspond to all the \gls{oos} lower triangle cumulative cash flows at time period $k$.

For convenience we substitute $j^*=k-i+1$ which corresponds to the latest development period for the given origin period $i$ and triangle calendar period $k$ so that, for example $X_{i,j^*}$ corresponds to a diagonal value.

To simplify the notation for conditional moments, we use subscripts on the expectation and variance operators to denote the information available up to the $j$th development period. Specifically, for any random variable $X$, \[E_{i,j}(X):=E(X|C_{i,1},C_{i,2},\ldots,C_{i,j}).\]

The chain ladder method places substantial emphasis on the ratios of successive cumulative cash flows. These empirical quantities, referred to as \textit{individual development ratios}, are defined as \[f_{i,j}:=\frac{C_{i,j+1}}{C_{i,j}},\] so that the identity $C_{i,j}=C_{i,1}\cdot\prod^{j-1}_{l=1}f_{i,l}$ holds true.

From a predictive perspective, the method seeks to estimate the development ratio parameters $f_j$ that generalise these ratios across accident periods ($i$) for each development period. Outstanding claim developments are then predicted by assuming that the proportional change in cumulative cash flows from one development period to the next is stable across accident periods.

This idea is formalised through the assumption: \[E_{i,j}(C_{i,j+1})=C_{i,j}\cdot f_j.\] This leads to the one-step projection definition \[\hat C_{i,j+1}=C_{i,j}\cdot f_{j},\] which can then be generalised to obtain the ultimate loss estimate \[\hat C_{i,J}^{(k)}=C_{i,j^*}\cdot \prod_{l=j^*}^{J-1} f_l.\] Here $\hat C_{i,j}^{(k)}$ denotes a projection made with all the information in $\Delta^k$ for accident period $i$ and development period $j$.

Considering the chain ladder as a predictive framework allows for various methods of obtaining parameter estimates for $f_j$. In practice they are often selected manually by choosing ad-hoc ratios or Bayesian style adjustments, allowing reserving actuaries to apply expert-informed judgement. Parameter estimation can also be framed as minimising a loss function, an approach common in machine learning. This idea has been fundamental to recent research in loss reserving and has led to models with better \gls{oos} performance than the more traditional methods such as the \gls{vwa} chain ladder \citep{ref:wuthrich2018neural, ref:AIinActSci, ref:CDRScore}.

\cite{ref:mack1994} showed how treating the development ratios of the chain ladder as model parameters can be used to derive the \gls{vwa} solution as optimal parameter estimates, by showing that it minimises a loss function. This helped make some of the implicit assumptions underlying the \gls{vwa} solution mathematically explicit. \cite{ref:mack1994} used the assumption \[C_{i,j+1}\approx C_{i,j}\cdot f_j,\] that is, the one-step predictions should predict the true values well.

This relationship between the one-step predictions and the true values is treated as probabilistic, that is, it has a (random) error component. One way to treat this error is to assume it is additive, so that
\begin{align}\label{eq:additiveError}
	C_{i,j+1}= \hat C_{i,j+1}+\epsilon_{i,j}=C_{i,j}\cdot f_{j}+\epsilon_{i,j},
\end{align}
where the $\epsilon_{i,j}$ values are independent random variables, with $E_{i,j}(\epsilon_{i,j})=0$. The error terms can be written as $C_{i,j+1}-C_{i,j}\cdot f_{j}= \epsilon_{i,j}.$ This is the error assumption made by \cite{ref:Mack1993,ref:mack1994}, and the one this paper challenges. Note that the zero-mean assumption is equivalent to \[E_{i,j}(C_{i,j+1})=C_{i,j}\cdot f_j.\] It follows that, under these assumptions, choosing $\boldsymbol{f}$ to minimise the sum of squared errors gives an unbiased estimator of the development ratios:
\begin{align}
	\sum^{J-1}_{j=1}\sum_{i=1}^{I-j} \epsilon_{i,j}^2 & = \sum^{J-1}_{j=1}\sum_{i=1}^{I-j}(C_{i,j+1}-C_{i,j}\cdot f_{j})^2.
\end{align}

When minimising this loss function, the different values of $f_j$ do not influence one another. Since each $f_j$ is only influenced by claim amounts in development period $j$ and individual development ratios $f_{j}$ and $f_{j+1}$, minimising the full objective reduces to minimising each inner sum over the accident periods separately. This corresponds to the assumption of \cite{ref:Mack1993} that the series of claims over accident periods are independent.

\begin{align}
	\sum_{i=1}^{I-j} \epsilon_{i,j}^2 & =\sum_{i=1}^{I-j}(C_{i,j+1}-C_{i,j}\cdot f_{j})^2=\sum_{i=1}^{I-j}C_{i,j}^2\cdot(f_{i,j}- f_{j})^2
\end{align}

By the Gauss-Markov theorem, minimising the unweighted sum of squared errors yields the best linear unbiased estimator only when the errors are uncorrelated and homoscedastic. However, \cite{ref:Mack1993} argued that, conditional on the claims observed to date, the variance of $C_{i,j+1}$ can be written as \[\text{Var}_{i,j}(C_{i,j+1})=C_{i,j}\cdot\alpha_j^2,\] with $\alpha_j$ constant across accident periods $i$. For a fixed development period $j$, this conditional variance is proportional to $C_{i,j}$ and therefore varies across accident periods $i$. The errors are thus heteroscedastic, and ordinary least squares is no longer optimal. Instead, \cite{ref:mack1994} used a weighted sum of squares, with weights equal to the precision of the errors (the reciprocal of the variance), $w_i=\frac{1}{C_{i,j}}$. The resulting loss function is displayed in Equation~\eqref{eq:VWMSE}.

\begin{align}\label{eq:VWMSE}
	\sum_{i=1}^{I-j}\frac{1}{C_{i,j}}\cdot \epsilon_{i,j}^2 & =\sum_{i=1}^{I-j}\frac 1 {C_{i,j}}\cdot (C_{i,j+1}-C_{i,j}\cdot f_{j})^2=\sum_{i=1}^{I-j}C_{i,j}\cdot(f_{i,j}- f_{j})^2
\end{align}

By taking the derivative with respect to $f_j$ and equating it to zero, it can be shown that the \gls{vwa} chain ladder solution minimises the loss function \citep{ref:mack1994}. That is, the minimiser is the weighted average development ratio, weighted by the cumulative cash flows as follows:
\begin{align}
	\hat f_j^{\text{VWA}}&=\frac{\sum^{I-j}_{i=1}  C_{i,j}\cdot f_{i,j}}{\sum^{I-j}_{i=1} C_{i,j}}\\
	&=\frac{\sum^{I-j}_{i=1} C_{i,j+1}}{\sum^{I-j}_{i=1}  C_{i,j}}.
\end{align}
The second equality holds whenever $C_{i,j}>0$ for every accident period entering the column-$j$ average. Where a cumulative claim amount is zero the individual development ratio is undefined and the observation is excluded, as described in Appendix~\ref{app:datasetsSelection}, so the two expressions no longer coincide. Throughout this paper we use the first form, the weighted average of the individual development ratios.

\section{A multiplicative error for chain ladder}\label{sec:MultErr}

This section derives the \gls{vwga} estimator and shows that it arises from three independent lines of reasoning. Section~\ref{sec:VWGADer1} obtains it as the minimiser of a volume-weighted squared log-error loss - the direct multiplicative-error analogue of Mack's additive derivation. Section~\ref{sec:VWGADer2} shows that the same estimator follows from requiring stability in successive ultimate loss projections. Section~\ref{sec:VWGADer3} recovers it once more as the maximum-likelihood estimate under a log-normal model for the individual development ratios. The first two derivations are distribution-free, in keeping with the framework of \cite{ref:Mack1993}; the third is parametric and included for additional motivation.

\subsection{The multiplicative-error assumption and the \texorpdfstring{\gls{vwga}}{VWGA} solution}\label{sec:VWGADer1}

As has already been discussed, the chain ladder method's predictive framework is built to incorporate the intrinsic multiplicative relationship of cumulative cash flows over the same accident years, yet the \gls{vwa} solution is based on an additive assumption of the error, as expressed in Equation~\eqref{eq:additiveError}. Concretely, this paper asks whether the relationship between successive cash flows is multiplicative in both the development ratios and the successive prediction error. That is: \[C_{i,j+1}=C_{i,j}\cdot f_{j}\cdot\epsilon_{i,j},\] or equivalently: \[\epsilon_{i,j}=\frac{C_{i,j+1}}{C_{i,j}\cdot f_{j}}.\]

This assumption leads to a new loss function, which can be derived by first noting that a perfect fit now corresponds to $\epsilon_{i,j}=1$, or equivalently $\log(\epsilon_{i,j})=0$. An effective way of handling this is by minimising the sum of the squared log-errors.
\begin{align}
	\sum^{I-j}_{i=1} \log^2(\epsilon_{i,j}) & =\sum^{I-j}_{i=1} \log^2\left(\frac{C_{i,j+1}}{C_{i,j}\cdot f_{j}}\right)=\sum^{I-j}_{i=1} \log^2\left(\frac{f_{i,j}}{f_{j}}\right) \\&=\sum^{I-j}_{i=1} (\log({f_{i,j}})-\log({f_{j}}))^2
\end{align}

A ratio-based loss such as the mean squared percentage error would also work, but it is asymmetric: over- and under-predictions of the same multiplicative size incur unequal penalties. This asymmetry is not a problem when using logs, because \[\left(\log\left(\frac a b\right)\right)^2=\left(-\log\left(\frac b a\right)\right)^2=\left(\log\left(\frac b a\right)\right)^2.\]

Minimising the sum of squared log-errors would lead to an unbiased model with $E_{i,j}[\log(f_{i,j})]=\log(f_j)$. Therefore, we assume that \[E_{i,j}(\log(C_{i,j+1}))=\log(C_{i,j}\cdot f_j),\] so that the model is unbiased on a logarithmic scale. This replaces the mean-scale assumption $E_{i,j}(C_{i,j+1})=C_{i,j}\cdot f_j$ made by \cite{ref:Mack1993} and shown in Section~\ref{sec:background}: by Jensen's inequality the two cannot hold simultaneously unless the error is degenerate. The bias that each choice induces on the other scale is examined in Section~\ref{sec:properties}.

To find the parameters giving the best linear unbiased estimate, we must determine whether the log-errors are heteroscedastic, which requires an assumption about the variance of the log-claims.

\cite{ref:Mack1993} argued, based on the \gls{vwa} chain ladder solution, that the variance of the individual development ratios should be $\text{Var}_{i,j}(f_{i,j})=\frac {\alpha_j^2} {C_{i,j}}$. Using a Taylor series expansion, it can be shown that when this variance is assumed and the model fits the training data relatively well - that is, when $C_{i,j+1} \approx C_{i,j} \cdot f_j$ - the variance of the errors is similar to its log-error counterpart. Appendix~\ref{app:VarianceDerivation} derives this result.

Thus, borrowing from \cite{ref:Mack1993}, we assume that the variance of the log-errors is \[\text{Var}_{i,j}\left(\log\left(\frac{f_{i,j}}{f_j}\right)\right)=\frac {\alpha_j^2} {f_j^2\cdot C_{i,j}}\propto\frac{1}{C_{i,j}},
\] the proportionality occurs because $f_j$ and $\alpha_j$ are constant across accident periods for a given $j$. This leads to the weighting: $w_i= C_{i,j}$, which can be used to obtain the final loss function as the volume-weighted sum of squared log-errors.

\begin{align}
	\sum^{I-j}_{i=1} C_{i,j}\cdot \log^2(\epsilon_{i,j}) & =\sum^{I-j}_{i=1} C_{i,j}\cdot(\log (C_{i,j+1})-\log (C_{i,j}\cdot f_{j}))^2 \\\label{eq:logLoss}&=\sum^{I-j}_{i=1} C_{i,j}\cdot\left(\log({f_{i,j}})-\log({f_{j}})\right)^2
\end{align}

Minimising this loss function with respect to the development ratios leads to the \gls{vwga} of individual development ratios. This is identical to the \gls{vwa} solution, except that a geometric mean is used instead of an arithmetic mean. The weights for \gls{vwga} warrant comment. Had the log-errors been homoscedastic, the minimiser would have been the unweighted geometric mean. It is the variance assumption of \cite{ref:Mack1993} that supplies the precision weights $C_{i,j}$ and thereby leads to the \gls{vwga}. The derivation of this result is given in Appendix~\ref{app:VWGADerivation}. The \gls{vwga} solution is given in Equations~\eqref{eq:geomMeanSolution1}--\eqref{eq:geomMeanSolution2}. The latter form is more useful in practice, since it uses logarithms to work with less extreme values and avoid numerical instability.

\begin{align}\label{eq:geomMeanSolution1}
	\hat f_j^{\text{VWGA}} & =\left(\prod^{I-j}_{i=1}\left(\frac{C_{i,j+1}}{C_{i,j}}\right)^{C_{i,j}}\right)^{\frac 1 {\sum^{I-j}_{i=1} {C_{i,j}}}} =\left(\prod^{I-j}_{i=1}f_{i,j}^{C_{i,j}}\right)^{\frac 1 {\sum^{I-j}_{i=1} {C_{i,j}}}} \\\label{eq:geomMeanSolution2}&=\exp\left[\frac{1}{\sum^{I-j}_{i=1} C_{i,j}}\sum^{I-j}_{i=1} C_{i,j}\cdot \log({f_{i,j}})\right]
\end{align}

\subsection{An alternative derivation from ultimate loss stability}\label{sec:VWGADer2}

The \gls{vwga} solution can also be motivated by targeting stable successive ultimate loss projections and assuming a multiplicative error. The solution therefore optimises a dual objective.

\cite{ref:merz2008CDR} showed how aggregating the \gls{cdr} can be used to estimate the one-year volatility of successive ultimate loss projections, where smaller values correspond to a more stable model that produces good reserve estimates. \cite{ref:CDRScore} built a loss function from it, using its properties for better ultimate loss projections. The \gls{cdr} assumes that successive ultimate loss projections should be similar, but since it considers the difference between them, it again assumes an additive relationship. It is natural to ask what the corresponding multiplicative assumption implies.

A multiplicative error between successive ultimate losses fits naturally, given the geometric form of the ultimate loss predictions. Using the notation introduced in Section~\ref{sec:background} this can be written as: \[\hat C_{i,J}^{(k+1)}= \hat C_{i,J}^{(k)}\cdot \xi_{i,j^*}=C_{i,j^*}\cdot\left(\prod^{J-1}_{l=j^*}f_l\right)\cdot\xi_{i,j^*}.\] This assumption implies that \[\xi_{i,j^*}=\frac{\hat C_{i,J}^{(k+1)}}{\hat C_{i,J}^{(k)}}=\frac{C_{i,j^*+1}}{C_{i,j^*}\cdot f_{j^*}}=\epsilon_{i,j^*},\] i.e. that this error is identical to the one assumed when deriving the \gls{vwga} solution. Because these errors are identical to those in Section~\ref{sec:VWGADer1}, the same assumptions apply, giving the same loss function and hence the \gls{vwga} solution. It follows that the \gls{vwga} solution optimises two desirable objectives for reserving models. That is, accurately predicting cumulative cash flows, and pursuing stable consecutive year-on-year ultimate loss projections. This has the potential to lead to better \gls{oos} performance, especially for ultimate loss projections, and motivates the empirical comparison in Section~\ref{sec:results}. Note that Section~\ref{sec:results} evaluates the accuracy of the resulting projections; a direct evaluation of the one-year stability itself, for example through the \gls{cdr}, is left for future work.

\subsection{Log-normal alternative derivation}\label{sec:VWGADer3}

For a third justification, we can derive the \gls{vwga} solution by assuming a log-normal distribution for the individual development ratios. Note that the other derivations were distribution-free.

Assuming the individual development ratios $f_{i,j}$ are themselves normally distributed recovers the \gls{vwa} solution, since by the well-known connection between least squares and normality its maximum-likelihood estimate is the least-squares solution. A major limitation of this assumption, however, is that the support of the normal distribution is unbounded, whereas the individual development ratios are strictly positive. Assuming instead that the individual development ratios are log-normally distributed, i.e. \[\log \left(f_{i,j}\right)\sim\text{N}\left(\log (f_j),\frac{\alpha_j^2}{C_{i,j}}\right),\] the maximum likelihood estimate of $f_j$ can be shown to equal the \gls{vwga} solution; for completeness this result is proven in Appendix~\ref{app:LNDer}. Note that this makes the same moment assumption as in Section~\ref{sec:VWGADer1}.

Interestingly, \cite{ref:hertig1985} suggested a similar approach for modelling the run-off pattern of reinsurance treaties, assuming the loss-quotient development ratios to be log-normally distributed. The main difference was that a constant variance was assumed across all underwriting years.

\section{Properties of the \texorpdfstring{\gls{vwga}}{VWGA} solution}\label{sec:properties}

The use of a geometric average pattern is not without precedent in the actuarial literature. \cite{ref:zehnwirth1997crm} mentions that ``the geometric mean of development factors is a more efficient estimate'' and \cite{ref:friedland2010} lists the simple geometric average among the standard averaging choices for development ratios. This paper contributes to the literature by introducing the claims volume weighting for the geometric average pattern.

The derivation of the \gls{vwga} in Appendix~\ref{app:VWGADerivation} assumes positive cumulative claim values throughout the triangle. Having zero triangle values is common in practice. To handle this, these ratios can simply be excluded from the pattern calculation.

It is a well-known result that the weighted geometric mean is less than or equal to the weighted arithmetic mean, with equality if and only if all the averaged values coincide. Applying this with weights $C_{i,j}$ to the individual development ratios gives $\hat f_j^{\text{VWGA}}\le \hat f_j^{\text{VWA}}$ for every $j$, with equality if and only if the ratios $f_{i,j}$ entering the column-$j$ average are all equal. Consequently, the \gls{vwga} ultimate loss predictions are less than or equal to those of the \gls{vwa}.

It is important to note that the \gls{vwga} method is unbiased on a logarithmic scale and biased with respect to the unit scale of individual development ratios $f_{i,j}$. This is because the natural logarithm, which is concave, is used in the loss function and through Jensen's inequality it can be shown that $\log(E_{i,j}(\epsilon_{i,j}))\ge E_{i,j}(\log(\epsilon_{i,j}))=0$. Conversely, the \gls{vwa} method is unbiased on the unit scale of individual development ratios $f_{i,j}$ and biased on a logarithmic scale. The choice between the methods therefore reflects competing assumptions about whether unbiasedness matters more on the logarithmic or the unit scale, rather than one method being more correct than the other.

\cite{ref:Mack1993} showed the volume-weighted chain ladder to be an unbiased estimate under certain assumptions. Even so, \cite{ref:CLBiasOld} found that the \gls{vwa} chain ladder exhibited a positive bias in simulation studies, producing estimates that systematically over-predict future losses. These findings were echoed by \cite{ref:CLbiasNew}. This suggests a problem with the fundamental assumptions of the volume-weighted chain ladder. Given that ultimate loss projections under \gls{vwga} are never larger than those of the \gls{vwa}, these assumptions may help address the presence of an observed upward bias. This idea is further explored in an empirical study in Section~\ref{sec:BiasMPE}.

Considering the individual development ratios on a logarithmic scale is attractive because they are strictly positive quantities and the distribution of observed development ratios is often right-skewed. The logarithmic transformation reduces this skewness and places the data on an unbounded scale where symmetric error assumptions are more plausible. The \gls{vwga} is therefore the more appropriate method for practitioners who judge the ratios to be best represented on that scale.

The idea of considering a multiplicative error and using a geometric average of the development ratios \citep{ref:murphy1994unbiased} or even the incremental claims \citep{ref:geometricCL} is not new in the actuarial literature. This paper differs from \cite{ref:murphy1994unbiased}, critiqued by \cite{ref:murphy1994unbiasedCritique}, in its theoretical justification and in allowing the error variances to vary rather than holding them constant. Our solution therefore carries volume weightings, selecting the ratios on a logarithmic scale while placing more emphasis on larger claims.

The logarithmic transformation underlying the \gls{vwga} solution also makes it more robust to outliers than the \gls{vwa} solution. Because the loss is defined on log ratios, a single unusually large development ratio has a damped influence on the estimate, which limits the severe over-predictions the \gls{vwa} solution is prone to when outliers are present. This robustness applies to large outlier ratios rather than small ones.

The multiplicative-error assumption also connects the \gls{vwga} solution to the log-normal reserving literature discussed in the introduction. Taking logarithms of the one-step model gives
\begin{equation}
	\log (C_{i,j+1}) = \log (C_{i,j}) + \log (f_j) + \log(\epsilon_{i,j}),
\end{equation}
which, for a fixed development period $j$, is a linear model for the log development ratios with the single parameter $\log (f_j)$:
\begin{equation}
	\log (f_{i,j}) = \log (f_j) + \log(\epsilon_{i,j}).
\end{equation}
The \gls{vwga} solution is the weighted least squares estimator of this model, with the precision weights $C_{i,j}$ derived in Section~\ref{sec:MultErr}. Under the log-normal assumption introduced in Section~\ref{sec:VWGADer3}, this becomes a weighted log-normal regression whose maximum-likelihood equivalence to the \gls{vwga} is established there. In this way the derivations of this paper connect the distribution-free approach of \cite{ref:Mack1993} with the parametric log-normal literature \citep{ref:kremer1982,ref:verrall1991}.

The value of the volume-weighted log loss extends beyond the chain ladder. Because it is defined on ratios, it admits clean and interpretable constraints on the loss function. The same loss can carry over to machine-learning and other model-based reserving methods that already train under a loss function. For practitioners, the change is small: the \gls{vwga} ratios directly substitute for the \gls{vwa} ratios and leave the actuary's workflow largely unchanged.

\section{Empirical validation}

\subsection{Evaluation metrics}

Section~\ref{sec:MultErr} gave theoretical reasons for preferring the \gls{vwga} ratios; this section tests empirically whether they improve on the \gls{vwa}. A quantitative comparison requires evaluation metrics. Because the Schedule P data include \gls{oos} values, we can measure how well each method predicts unseen cash flows and ultimate loss projections.

In reserving, accurate ultimate loss projections are often the primary objective, so the first metric is the \gls{rmspe}, which measures how closely projected ultimates match the true values via squared percentage errors. Being unit free, it can be aggregated across datasets with a simple mean.

\begin{equation}
	\text{RMSPE} = \sqrt{\frac{1}{I-1}\sum^I_{i = 2} \left(\frac{\hat C^{(J)}_{i,J}-C_{i,J}} {C_{i,J}}\right)^2}
\end{equation}

We also examine the individual cash flow projections (the lower triangle), where offsetting errors do not cancel as they can at the aggregate level - the actual versus expected view \citep{ref:wuthrich2018neural,ref:CDRScore}. The \gls{maave} measures the discrepancy between actual and expected incremental values, and dividing it by the corresponding ultimate loss gives the unit-free \gls{mapave}.

\begin{align}
	\text{MAAvE}  & = \frac{1}{|\Delta^{-J}|}\sum_{(i,j):C_{i,j}\in\Delta^{-J}}\left|\hat X^{(J)}_{i,j}-X_{i,j} \right|                     \\
	\text{MAPAvE} & = \frac{1}{|\Delta^{-J}|}\sum_{(i,j):C_{i,j}\in\Delta^{-J}}\frac 1 {C_{i,J}}\cdot \left|\hat X^{(J)}_{i,j}-X_{i,j} \right|
\end{align}

To measure systematic bias, the \gls{mpe} of the ultimate loss projections is calculated as follows:
\begin{equation}
	\text{MPE} = \frac{1}{I-1}\sum^I_{i = 2} \frac{C_{i,J}-\hat C^{(J)}_{i,J}} {C_{i,J}}.
\end{equation}
Note that negative values correspond to over-predictions and positive values to under-predictions. Because the sign obscures the size of the bias when methods err in opposite directions, we also report the magnitude of a dataset's mean percentage error, denoted $|\text{MPE}|$:
\begin{equation}\label{eq:absMPE}
	|\text{MPE}| = \left|\frac{1}{I-1}\sum^I_{i = 2} \frac{C_{i,J}-\hat C^{(J)}_{i,J}} {C_{i,J}}\right|.
\end{equation}
This is not a mean absolute percentage error: the absolute value is taken \emph{after} averaging over the accident years, so over- and under-predictions within a dataset still cancel and only the direction of the net bias that survives them is discarded. Averaging the absolute percentage errors instead would measure accuracy, which is what the \gls{mapave} already does.

To compare the two methods on an aggregate level over the same datasets, we apply a paired $t$-test to the per-dataset differences in each metric, assuming those differences are normally distributed.

Let $d_m = E_m^\text{VWGA} - E_m^\text{VWA}$ denote the difference in error metrics for the $m$th of the $n$ datasets, one per company and line of business. The $t$-statistic is defined in Equations~\eqref{eq:t-stat1}--\eqref{eq:t-stat3}. It is approximately a $t$-random variable with $n-1$ degrees of freedom, yielding a $p$-value. We test the null hypothesis that the mean error metric is equal for the two methods against the one-sided alternative that it is smaller for the \gls{vwga}; a large negative $t$-statistic supports the alternative.
\begin{align}\label{eq:t-stat1}
	\bar d        & = \frac{1}{n} \sum_{m=1}^{n} d_m              \\
	S_d^2         & = \frac 1 {n-1}\sum_{m=1}^{n}(d_m - \bar d)^2 \\
	\text{SE}_d   & = \frac {S_d} {\sqrt {n}}\\\label{eq:t-stat3}
	t\text{-stat} & = \frac{\bar d}{\text{SE}_d }
\end{align}

Note that the 362 datasets are drawn from 207 companies, since a company writing several lines contributes one dataset per line, so the differences $d_m$ are not fully independent and the nominal degrees of freedom are overstated. Furthermore, the paid and incurred tests are run on the same companies over correlated triangles, so the two reported $p$-values should not be read as independent confirmations of one another.

\subsection{Results}\label{sec:results}

362 of the 772 datasets compiled by \cite{ref:ScheduleP} were selected. Both the selection criteria and the list of selected datasets are given in Appendix~\ref{app:datasetsSelection}.
Each evaluation metric is presented in turn. The Schedule P datasets were last updated in December 2025 and differ from the originally published data (September 2011).

\subsubsection{Root mean squared percentage error (\texorpdfstring{\gls{rmspe}}{RMSPE})}

First, consider the companies' \gls{rmspe}, which is summarised in Table~\ref{tab:RMSPE}. The \gls{vwga} pattern scores better than the \gls{vwa} on average for every line of business, across both paid and incurred projections, with the improvement most pronounced on paid data. The largest gains are in product liability and other liability, lines that tend to carry more outliers - consistent with the \gls{vwga}'s log transformation, which is more robust to large disparities between consecutive claim sizes.
For paid, the paired-difference $t$-statistic of the \gls{rmspe} is $-3.74$ ($\text{SE}_d = 0.0077$), with a $p$-value of 0.0001. For incurred, it is $-5.38$ ($\text{SE}_d = 0.0052$), with a $p$-value much less than 0.0001. Incurred has a more extreme $t$-statistic even though it displays less improvement, which can be attributed to its smaller standard error. This confirms the observations from this section: the \gls{vwga} outperforms the \gls{vwa} for both paid and incurred projections under the \gls{rmspe}.

\begin{table}[!htbp]
	\centering
	\renewcommand{\arraystretch}{1.2}
	\setlength{\tabcolsep}{2pt}
	\begin{tabular}{l c c c c c c c}
		\toprule
		\rowcolor{gray!15}
		                    & \textbf{ComAuto} & \textbf{MedMal} & \textbf{PPAuto} & \textbf{WorkComp} & \textbf{ProdLia} & \textbf{OthLia} & \textbf{All} \\
		\midrule
		{\textbf{Incurred}} &                  &                 &                 &                   &                  &                 &              \\
		\gls{vwa}                 & 0.2205           & 0.3745          & 0.1156          & 0.2500            & 0.7088           & 0.4477          & 0.2763       \\
		\gls{vwga}                & 0.2038           & 0.3658          & 0.1088          & 0.2439            & 0.5906           & 0.3833          & 0.2485       \\
		\midrule
		{\textbf{Paid}} &                  &                 &                 &                   &                  &                 &                  \\
		\gls{vwa}                 & 0.2322           & 0.3713          & 0.0843          & 0.2119            & 0.5495           & 0.3718          & 0.2395       \\
		\gls{vwga}                & 0.2095           & 0.3477          & 0.0809          & 0.2068            & 0.4405           & 0.3055          & 0.2107       \\
		\bottomrule
	\end{tabular}
	\caption{Out-of-sample \gls{rmspe} for different chain ladder methods.}
	\label{tab:RMSPE}
\end{table}

\begin{figure}[!htbp]
	\centering
	\includegraphics[width=0.75\textwidth]{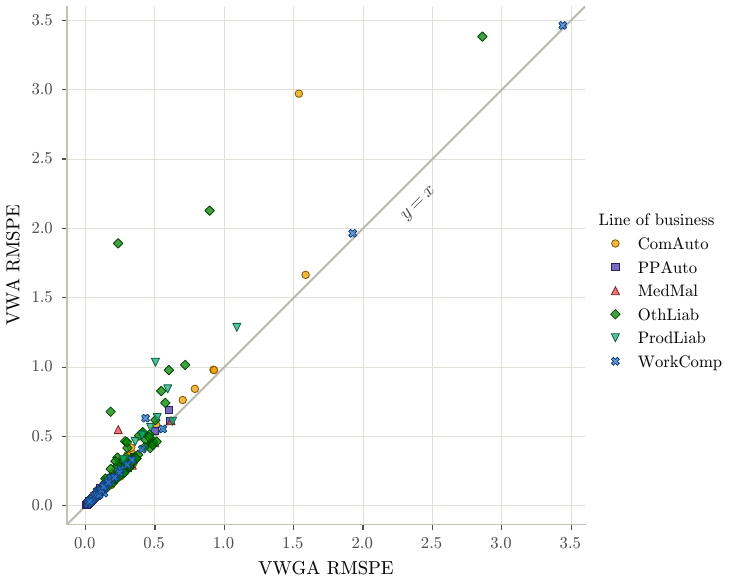}
	\caption{Scatter plot of \gls{vwa}'s \gls{rmspe} plotted against \gls{vwga}'s \gls{rmspe} values, for each \textbf{paid} dataset.}
	\label{fig:RMSPEPaid}
\end{figure}

Figure~\ref{fig:RMSPEPaid} plots the two methods' \gls{rmspe} against each other for every paid dataset. The identity line ($y=x$) separates the metrics visually: points above the line represent datasets where the \gls{vwga} performs better, points below where the \gls{vwa} does. It shows the \gls{vwga} outperforming the \gls{vwa} in most cases. The gap is widest for datasets where both methods predict the \gls{oos} data poorly (\gls{rmspe} above 40\%). For many of these cases the \gls{vwa} method was too sensitive to outliers, which led to severe over-predictions. This implies that the \gls{vwga} limits the extent of over-predictions when outliers are present. Compared with simply excluding outliers, this approach still retains some of their information. These poorly-fitting datasets account for most of the aggregate improvement, and for paid data for almost all of it; Appendix~\ref{app:stratification} stratifies the results by accuracy to show where the gain arises.

\begin{figure}[!htbp]
	\centering
	\includegraphics[width=0.75\textwidth]{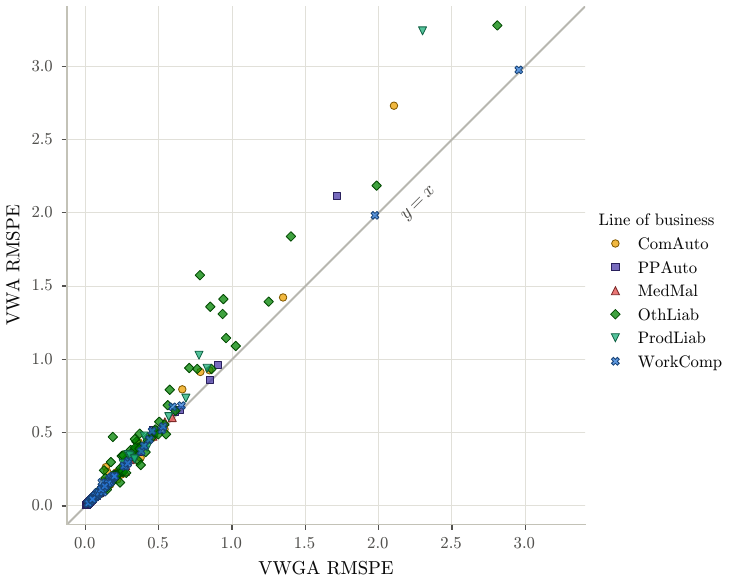}
	\caption{Scatter plot of \gls{vwa}'s \gls{rmspe} plotted against \gls{vwga}'s \gls{rmspe} values, for each \textbf{incurred} dataset.}
	\label{fig:RMSPEIncurred}
\end{figure}

Figures~\ref{fig:RMSPEPaid} and \ref{fig:RMSPEIncurred} show that the results for workers compensation are comparable across the two chain ladder methods. On closer inspection, these points lie slightly above the identity line ($y=x$), indicating a small improvement for the \gls{vwga}. This result is consistent when aggregating across all companies as shown in Table~\ref{tab:RMSPE}; the \gls{vwga} method performed at least as well as \gls{vwa} in every scenario.

Across both commercial and private automotive insurance classes, the \gls{vwga} method marginally outperforms the \gls{vwa} approach in most scenarios and performs substantially better in the extreme cases where the \gls{rmspe} is large. This improved performance can be attributed to the greater robustness of \gls{vwga} where there are outliers in the dataset.

For medical malpractice, the limited number of observations prevents any generalisable conclusions.

The \gls{vwga} has a particularly strong advantage in product liability and other liability, where it generally yields substantial reductions in \gls{rmspe}. This highlights an opportunity to use \gls{vwga} instead of \gls{vwa} in liability-type lines, where greater variability and extreme observations are present.

\subsubsection{Mean absolute percentage actual versus expected (\texorpdfstring{\gls{mapave}}{MAPAvE})}

The \gls{mapave} results are summarised in Table~\ref{tab:MAPAvE}. Similar to the \gls{rmspe} results, the \gls{vwga} method consistently produces lower prediction errors than the \gls{vwa} method across most lines of business and for both paid and incurred projections. This shows the \gls{vwga} improves prediction accuracy at a granular cash flow level as well as at the aggregate ultimate level.

For incurred data, the average \gls{mapave} across all datasets decreases from 5.83\% under \gls{vwa} to 5.55\% under \gls{vwga}. For paid data, the average error decreases from 4.12\% to 3.79\%. Although these absolute improvements appear modest, they are consistent across nearly all lines of business. As was observed for \gls{rmspe}, the largest reductions occur in product liability and other liability business, where the greater variability in claims emergence gives rise to extreme observations.

The paired-difference $t$-test further supports these findings. For paid data, the test yields a $t$-statistic of $-4.49$ ($\text{SE}_d=0.00072$) with a $p$-value substantially below conventional significance levels. For incurred data, the $t$-statistic is $-5.20$ ($\text{SE}_d=0.00054$), again with a $p$-value effectively equal to zero. The reduction in \gls{mapave} achieved by the \gls{vwga} is therefore unlikely to be due to chance. Taken together with the \gls{rmspe} results, this indicates a meaningful improvement in predictive performance over the \gls{vwa} for both paid and incurred projections.

\begin{table}[!htbp]
	\centering
	\renewcommand{\arraystretch}{1.2}
	\setlength{\tabcolsep}{2pt}

	\begin{tabular}{l c c c c c c c}
		\toprule
		\rowcolor{gray!15}
		                    & \textbf{ComAuto} & \textbf{MedMal} & \textbf{PPAuto} & \textbf{WorkComp} & \textbf{ProdLia} & \textbf{OthLia} & \textbf{All} \\
		\midrule
		{\textbf{Incurred}} &                  &                 &                 &                   &                  &                 &              \\
		\gls{vwa}                 & 0.0425           & 0.0989          & 0.0208          & 0.0480            & 0.1473           & 0.1039          & 0.0583       \\
		\gls{vwga}                & 0.0405           & 0.0984          & 0.0200          & 0.0473            & 0.1360           & 0.0978          & 0.0555       \\
		\midrule
		{\textbf{Paid}} &                  &                 &                 &                   &                  &                 &                  \\
		\gls{vwa}                 & 0.0412           & 0.0654          & 0.0149          & 0.0308            & 0.0991           & 0.0653          & 0.0412       \\
		\gls{vwga}                & 0.0383           & 0.0644          & 0.0146          & 0.0304            & 0.0859           & 0.0579          & 0.0379       \\
		\bottomrule
	\end{tabular}
	\caption{Out-of-sample \gls{mapave} for different chain ladder methods.}
	\label{tab:MAPAvE}
\end{table}

\begin{figure}[!htbp]
	\centering
	\includegraphics[width=0.75\textwidth]{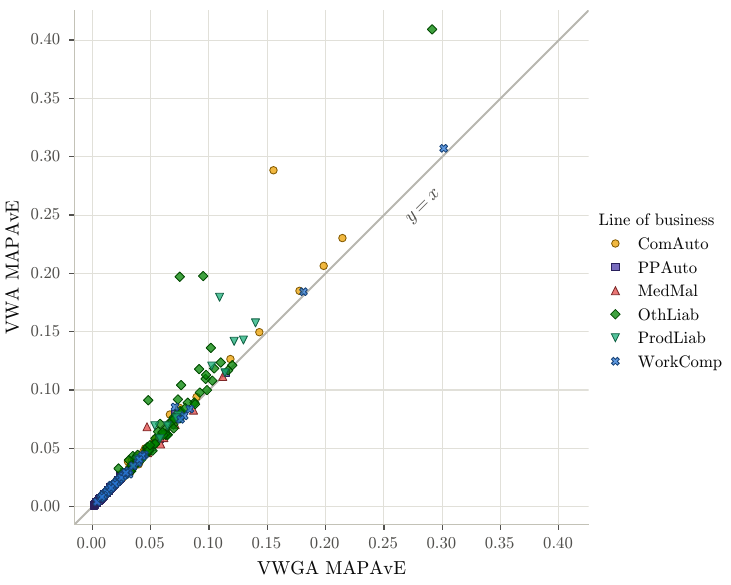}
	\caption{Scatter plot of \gls{vwa}'s \gls{mapave} plotted against \gls{vwga}'s \gls{mapave} values, for each \textbf{paid} dataset.}
	\label{fig:MAPAvEPaid}
\end{figure}

\begin{figure}[!htbp]
	\centering
	\includegraphics[width=0.75\textwidth]{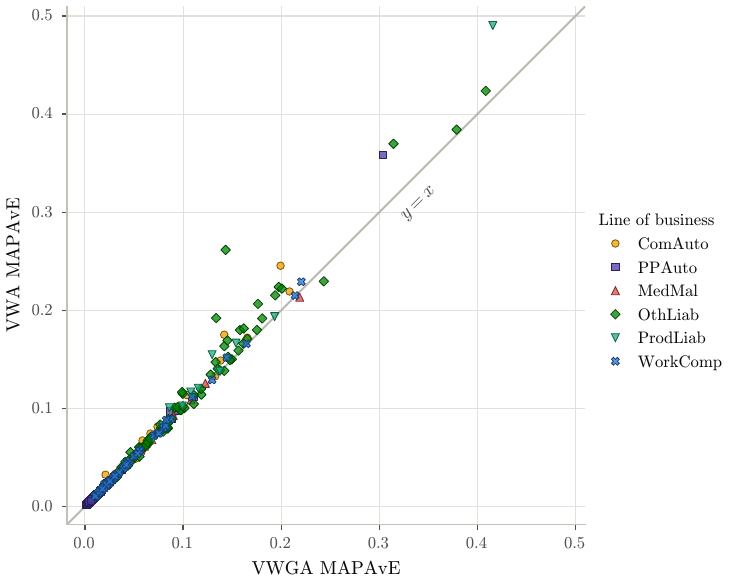}
	\caption{Scatter plot of \gls{vwa}'s \gls{mapave} plotted against \gls{vwga}'s \gls{mapave} values, for each \textbf{incurred} dataset.}
	\label{fig:MAPAvEIncurred}
\end{figure}

Figures~\ref{fig:MAPAvEPaid} and~\ref{fig:MAPAvEIncurred} provide a visual comparison of the two methods across individual datasets. As with the \gls{rmspe} scatterplots, points lying above the identity line ($y=x$) indicate superior performance by the \gls{vwga} method. Most observations fall above this line, demonstrating that \gls{vwga} achieves lower \gls{mapave} values for the majority of datasets.

The scatterplots further reveal that the improvement associated with \gls{vwga} becomes more noticeable as the prediction error increases. For datasets where both methods exhibit small \gls{mapave} values, the two approaches perform more similarly. However, for datasets with relatively poor predictive performance, \gls{vwga} often produces substantially smaller errors than \gls{vwa}. This pattern is particularly evident in liability-type business, where a small number of atypical development observations can materially influence development ratios. The logarithmic transformation underlying \gls{vwga} appears to mitigate the influence of these observations, reducing the tendency for extreme over-projections while still retaining information from the underlying data.

\FloatBarrier

\subsubsection{Mean percentage error (\texorpdfstring{\gls{mpe}}{MPE})}\label{sec:BiasMPE}

Having established that the \gls{vwga} predicts more accurately, we next ask whether the two methods differ in bias. Each method's bias was tested by calculating the \gls{mpe} of the ultimate losses. These means are given in Table~\ref{tab:MPE}, which records the direction of the average error, and in Table~\ref{tab:absMPE}, which records its size. Aggregated over the insurance types, the \gls{vwga} solution leads to less bias for both paid and incurred data.

The geometric mean is always less than or equal to the arithmetic mean. It follows that all projections made by the \gls{vwga} are less than or equal to those made by the \gls{vwa}. Bias is therefore important for actuaries to consider before implementing the \gls{vwga}: if it tends to under-predict ultimate losses, it might be less suited for reserving applications.

\begin{table}[!htbp]
	\centering
	\renewcommand{\arraystretch}{1.2}
	\setlength{\tabcolsep}{2pt}

	\begin{tabular}{l c c c c c c c}
		\toprule
		\rowcolor{gray!15}
		                                        & \textbf{ComAuto} & \textbf{MedMal}   & \textbf{PPAuto} & \textbf{WorkComp} & \textbf{ProdLia} & \textbf{OthLia}   & \textbf{All} \\
		\midrule
		{\textbf{Incurred}} &                  &                   &                 &                   &                  &                   & \\
		\gls{vwa}                 & -0.0542          & -0.1405           & -0.0468         & -0.0952           & -0.3073          & -0.1601           & -0.0971 \\
		\gls{vwga}                & -0.0397          & -0.1224           & -0.0416         & -0.0868           & -0.2180          & -0.1057           & -0.0731 \\
		\midrule
		{\textbf{Paid}}     &                  &                   &                 &                   &                  &                   & \\
		\gls{vwa}                 & -0.0419          & \phantom{-}0.1613 & -0.0173         & -0.0469           & -0.1300          & -0.0374           & -0.0338 \\
		\gls{vwga}                & -0.0208          & \phantom{-}0.1985 & -0.0143         & -0.0405           & -0.0439          & \phantom{-}0.0215 & -0.0076 \\
		\bottomrule
	\end{tabular}

	\caption{Out-of-sample \gls{mpe} for different chain ladder methods.}
	\label{tab:MPE}
\end{table}

For a best-estimate, unbiased projection we expect average prediction errors close to zero. We do not observe this in \gls{vwa}. While the sign of the \gls{mpe} is close to balanced across the paid datasets (negative in 191 of 362), the mean \gls{mpe} is $-3.4\%$, so the over-predictions must outweigh the under-predictions in magnitude. That imbalance leaves the mean \gls{mpe} negative in five of the six paid lines of business and in every incurred line, and it is more pronounced on incurred data, where the \gls{mpe} is negative in 251 of 362 datasets and the mean \gls{mpe} is $-9.7\%$. The \gls{vwa} is therefore not unbiased in this study, which is in line with findings by \cite{ref:CLbiasNew} as discussed in the introduction.

Because the \gls{vwga} projections are never larger than those of the \gls{vwa}, its \gls{mpe} is the larger of the two in each of the 362 datasets on both triangles, so a paired test on the signed difference would only restate that inequality. To compare the size of the bias rather than its direction we therefore average each dataset's $|\text{MPE}|$ as defined in Equation~\eqref{eq:absMPE}; those means are given in Table~\ref{tab:absMPE}.

\begin{table}[!htbp]
	\centering
	\renewcommand{\arraystretch}{1.2}
	\setlength{\tabcolsep}{2pt}

	\begin{tabular}{l c c c c c c c}
		\toprule
		\rowcolor{gray!15}
		                    & \textbf{ComAuto} & \textbf{MedMal} & \textbf{PPAuto} & \textbf{WorkComp} & \textbf{ProdLia} & \textbf{OthLia} & \textbf{All} \\
		\midrule
		{\textbf{Incurred}} &                  &                 &                 &                   &                  &                 &              \\
		\gls{vwa}                 & 0.0922           & 0.1890          & 0.0581          & 0.1213            & 0.3896           & 0.2047          & 0.1295       \\
		\gls{vwga}                & 0.0851           & 0.1791          & 0.0539          & 0.1166            & 0.3198           & 0.1682          & 0.1138       \\
		\midrule
		{\textbf{Paid}} &                  &                 &                 &                   &                  &                 &                  \\
		\gls{vwa}                 & 0.0999           & 0.1983          & 0.0368          & 0.0853            & 0.2515           & 0.1572          & 0.1029       \\
		\gls{vwga}                & 0.0907           & 0.1985          & 0.0358          & 0.0820            & 0.1828           & 0.1258          & 0.0893       \\
		\bottomrule
	\end{tabular}

	\caption{Out-of-sample $|\text{MPE}|$, the size of the bias, for different chain ladder methods.}
	\label{tab:absMPE}
\end{table}

The \gls{vwga} carries the smaller mean $|\text{MPE}|$ in every line of business and on both triangles, with one exception: paid medical malpractice, where the two methods are effectively tied at 19.83\% against 19.85\%. Pooled over all datasets the mean $|\text{MPE}|$ falls from 10.29\% to 8.93\% on paid data and from 12.95\% to 11.38\% on incurred data. The paired-difference $t$-test on $|\text{MPE}|$ yields a $t$-statistic of $-3.25$ ($\text{SE}_d=0.0042$) for paid data and $-5.09$ ($\text{SE}_d=0.0031$) for incurred data, so the reduction in the size of the bias is significant on both triangles. That reduction is concentrated rather than spread evenly: the ten paid datasets whose \gls{vwa} $|\text{MPE}|$ exceeds 50\% account for 65\% of the total paid reduction, while across the 188 paid datasets whose \gls{vwa} $|\text{MPE}|$ is already below 5\% the \gls{vwga} is marginally worse in aggregate. The bias improvement therefore follows the same pattern as the accuracy improvement of Appendix~\ref{app:stratification}.

For other liability, the \gls{vwga} under-predicts paid data slightly where the \gls{vwa} still over-predicts. Medical malpractice is the one line whose paid bias the \gls{vwga} does not reduce, which might be a cause for concern or a sign of a strong claims inflation trend in the data. With only eight datasets available, however, the result may not be representative, so we do not pursue it further.

\begin{figure}[!htbp]
	\centering
	\includegraphics[width=0.75\textwidth]{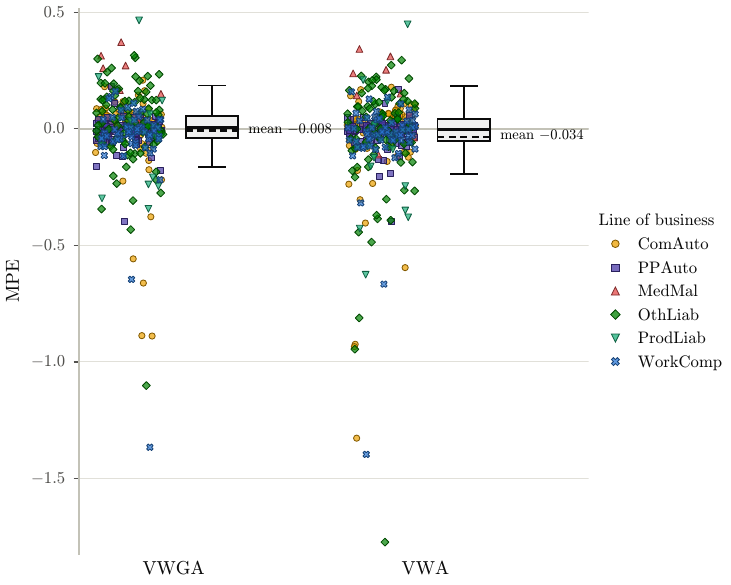}
	\caption{Box plot of \gls{vwa} and \gls{vwga}'s \gls{mpe} values, for each \textbf{paid} dataset. In each box the solid line is the median and the dashed line the mean.}
	\label{fig:MPEPaid}
\end{figure}

\begin{figure}[!htbp]
	\centering
	\includegraphics[width=0.75\textwidth]{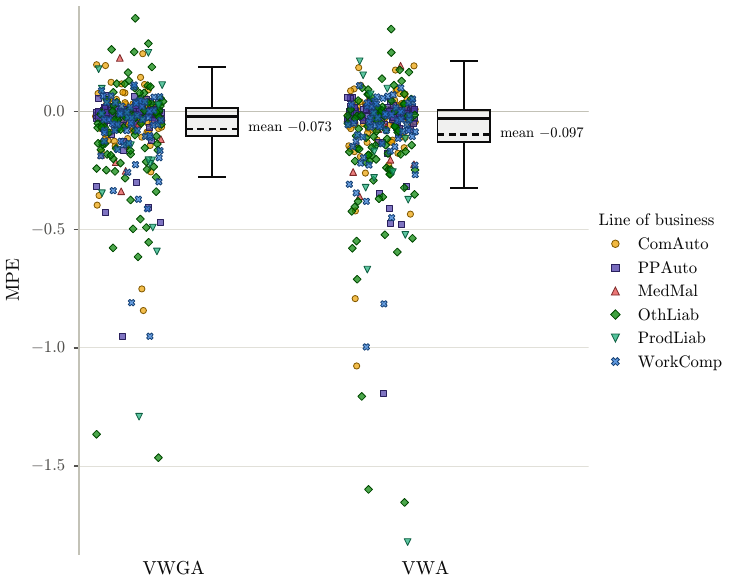}
	\caption{Box plot of \gls{vwa} and \gls{vwga}'s \gls{mpe} values, for each \textbf{incurred} dataset. In each box the solid line is the median and the dashed line the mean.}
	\label{fig:MPEIncurred}
\end{figure}

Figures~\ref{fig:MPEPaid} and~\ref{fig:MPEIncurred} show the two distributions. As expected, the \gls{vwga} predictions consistently have a larger \gls{mpe} than the \gls{vwa} predictions. The \gls{vwga} solution often reduced the over-predictions, which is likely due to its robustness in handling outliers. The dashed means in the figures make the shift explicit: the mean \gls{mpe} moves from $-3.4\%$ under the \gls{vwa} to $-0.8\%$ under the \gls{vwga} on paid data, and from $-9.7\%$ to $-7.3\%$ on incurred data.

\FloatBarrier

\section{Conclusion}

This paper examined whether aligning the error structure of the chain ladder with its multiplicative form is beneficial. The chain ladder predicts outstanding claims through successive multiplicative development ratios, yet the standard \gls{vwa} solution rests on an additive error assumption \citep{ref:Mack1993,ref:mack1994}. Starting from the multiplicative error and minimising a volume-weighted squared log-error loss in the spirit of Mack's distribution-free framework leads directly to the \gls{vwga} estimator. The \gls{vwga} is identical to the \gls{vwa} except that a volume-weighted geometric mean replaces the arithmetic mean of the individual development ratios, so both the error structure and the estimator match the multiplicative form the method already assumes.

The \gls{vwga} arises from three independent lines of reasoning, developed in Section~\ref{sec:MultErr}. It is the minimiser of a volume-weighted squared log-error loss under Mack's distribution-free assumptions; the estimator implied by requiring stability in successive ultimate loss projections; and the maximum-likelihood estimate under a log-normal model for the individual development ratios. The convergence of a distribution-free loss minimisation, a reserving-stability requirement, and a parametric likelihood leading to the same estimator provides better validation that no single derivation could offer.

The \gls{vwga} also preserves the practical character of the chain ladder. It has a closed form and a computational cost comparable to the \gls{vwa}'s, so the \gls{vwga} ratios substitute directly for the \gls{vwa} ratios, leaving the actuary's workflow and the flexibility of the chain ladder framework intact.

Beyond these theoretical arguments, the empirical study of Section~\ref{sec:results} supports the \gls{vwga}. Across 362 Schedule P company-line datasets, each with a paid and an incurred triangle spanning six lines of business, the \gls{vwga} matched or improved on the \gls{vwa}'s \gls{oos} predictions. It scored better on \gls{rmspe} and \gls{mapave} across nearly all lines and for both paid and incurred projections, and in aggregate it never scored worse on \gls{rmspe}. Paired $t$-tests confirmed that these improvements are statistically significant. The largest gains occurred in the liability lines - product liability and other liability - which tend to carry more outliers, consistent with the robustness the logarithmic transformation affords. The gain is uneven, however: as Appendix~\ref{app:stratification} shows, it is concentrated in the datasets that both methods predict poorly, and on paid triangles the two methods are difficult to distinguish once those datasets are set aside. The \gls{vwga} is therefore most valuable on triangles containing extreme individual development ratios, and close to neutral elsewhere.

Part of this improvement comes from a reduction in bias, examined in Section~\ref{sec:BiasMPE}. Because the weighted geometric mean is always less than or equal to the weighted arithmetic mean, the \gls{vwga} ultimate projections are never larger than those of the \gls{vwa} (Section~\ref{sec:properties}). The \gls{vwa}'s \gls{mpe} is negative in every incurred line of business and in five of the six paid lines, indicating a systematic tendency to over-predict, in line with the findings of \cite{ref:CLbiasNew}. The \gls{vwga} reduces this over-prediction bias, and a paired $t$-test confirms the reduction when the size of the bias is compared rather than its direction.

Taken together, these results suggest that the multiplicative-error assumption and the volume-weighted log loss are valuable for the chain ladder in practice and, more broadly, as a loss function for other reserving models, including machine-learning-based reserving. Because the log loss is defined on ratios, it transfers directly to modern methods that already train under a loss function (Section~\ref{sec:properties}).

Several limitations temper these findings and mark the next steps. The empirical study tested the accuracy of the \gls{vwga} projections, not the one-year stability its derivation in Section~\ref{sec:MultErr} implies, which a \gls{cdr}-based evaluation would measure. A natural extension is the volume-weighted log loss and the underlying multiplicative-error assumption applied to machine-learning-based reserving methods. The \gls{vwga} is also unbiased only on the logarithmic scale, and Section~\ref{sec:properties} sets out the trade-off this creates against the \gls{vwa}'s unbiasedness on the unit scale. Finally, the medical malpractice results rest on only eight datasets and are not generalisable, and the \gls{vwga} is more likely than the \gls{vwa} to under-predict, so its bias behaviour should be considered before it is adopted for a given portfolio.

\clearpage
\printnoidxglossary[type=\acronymtype,title={List of acronyms}]

\appendix
\newpage
\appendixpage

\section{Log-error variance derivation}\label{app:VarianceDerivation}

To motivate the variance assumption of the log-errors, we start from the assumptions of \cite{ref:Mack1993}, i.e. assuming that the variance of a cumulative cash flow given all the origin period's preceding cash flows is proportional to the previous cash flow, \[\text{Var}_{i,j}\left(C_{i,j+1}\right)=C_{i,j}\cdot\alpha^2_j.\] Next, consider the first-order Taylor series approximation \[\log(x)=\log(1+x-1)\approx x-1,\] which holds when $x\approx1$.

\begin{align}
	\text{Var}_{i,j}\left(\log(\epsilon_{i,j})\right) & = \text{Var}_{i,j}\left(\log\left(\frac{C_{i,j+1}}{C_{i,j}\cdot f_j}\right)\right) \\
	                                                  & \approx \text{Var}_{i,j}\left(\frac{C_{i,j+1}}{C_{i,j}\cdot f_j}-1\right)         \\
	                                                  & = \text{Var}_{i,j}\left(\frac{C_{i,j+1}}{C_{i,j}\cdot f_j}\right)                 \\
	                                                  & =\frac{\alpha^2_j}{C_{i,j}\cdot f_j^2}                                            \\
	                                                  & \propto\frac{1}{C_{i,j}}
\end{align}
This approximation holds when $\frac{C_{i,j+1}}{C_{i,j}\cdot f_j}\approx1$, that is, when the model fits the observed data well, which is a reasonable assumption when selecting the weighting. It follows that weighting the squared log errors by the cumulative claim amount, $C_{i,j}$, corresponds to the precision weighting.

\section{Geometric mean solution workings} \label{app:VWGADerivation}

Consider the terms of the volume-weighted sum of squared log errors that involve $f_j$, for $j\in\{1,2,\dots,J-1\}$.

\begin{equation}
	L_j =\sum^{I-j}_{i=1} C_{i,j}\cdot\left(\log({f_{i,j}})-\log({f_{j}})\right)^2+\text{Constant}
\end{equation}
Taking the derivative with respect to $f_j$ yields \[\frac{\partial L_j} {\partial f_j} =-2\cdot f_j^{-1}\cdot\sum^{I-j}_{i=1} C_{i,j}\cdot \left(\log({f_{i,j}})-\log({f_{j}})\right).\]

After equating to zero we obtain\[
	0=-\log(\hat{f}_{j}^{\text{VWGA}})\cdot\sum^{I-j}_{i=1} C_{i,j} + \sum^{I-j}_{i=1} C_{i,j}\cdot\log{f_{i,j}}.
\]
This simplifies to the \gls{vwga}:
\begin{align}
	{\hat{f}_{j}^{\text{VWGA}}} & =\exp\left[\frac{1}{\sum^{I-j}_{i=1} C_{i,j}}\sum^{I-j}_{i=1} C_{i,j}\cdot \log({f_{i,j}})\right] \\&=\left(\prod^{I-j}_{i=1} \left(\frac{C_{i,j+1}}{C_{i,j}}\right)^{C_{i,j}}\right)^{\frac 1 {\sum^{I-j}_{i=1}  {C_{i,j}}}}.
\end{align}

To confirm this is a minimum, we apply the second derivative test. Taking the second derivative with respect to $f_j$ gives \[\frac{\partial^2 L_j} {\partial f_j^2} = 2\cdot f_j^{-2}\cdot\sum^{I-j}_{i=1} C_{i,j}\cdot\log({f_{i,j}})-2\cdot \frac{\log({f_{j}})}{f_j^{2}}\cdot\sum^{I-j}_{i=1} C_{i,j}+2\cdot f_j^{-2}\cdot\sum^{I-j}_{i=1} C_{i,j}.\]

To find the sign of this function, we divide by $2\cdot f_j^{-2}$, since $f_j>0$ for all $j\in\{1,2,\dots,J-1\}$. Furthermore, plugging in $f_j=\hat{f}_{j}^{\text{VWGA}}$ yields
\begin{align}\nonumber
	\left.\frac{\partial^2 L_j} {\partial f_j^2}\right|_ {f_j = \hat{f}_{j}^{\text{VWGA}} } & \propto\sum^{I-j}_{i=1} C_{i,j}\cdot\log({f_{i,j}})-\log({\hat{f}_j^{\text{VWGA}}})\cdot\sum^{I-j}_{i=1} C_{i,j}+\sum^{I-j}_{i=1} C_{i,j}                                           \\\nonumber
	                                                                                        & =\sum^{I-j}_{i=1} C_{i,j}\cdot\log({f_{i,j}})-\frac{1}{\sum^{I-j}_{i=1} C_{i,j}}\sum^{I-j}_{i=1} C_{i,j}\cdot \log({f_{i,j}})\cdot\sum^{I-j}_{i=1} C_{i,j}+\sum^{I-j}_{i=1} C_{i,j} \\\nonumber
	                                                                                        & =\sum^{I-j}_{i=1} C_{i,j}\cdot\log({f_{i,j}})-\sum^{I-j}_{i=1} C_{i,j}\cdot \log({f_{i,j}})+\sum^{I-j}_{i=1} C_{i,j}                                                                \\\nonumber
	                                                                                        & =\sum^{I-j}_{i=1} C_{i,j}>0.
\end{align}

When working with the individual development ratios, it is assumed that all cumulative claims are positive, that is, $C_{i,j}>0$ for the entire triangle. It follows that the second derivative is always positive. The function is therefore at a minimum at the \gls{vwga} solution. This is further discussed in Section~\ref{sec:properties}.

\section{Log-normal derivation} \label{app:LNDer}

Assume that \[\log \left(f_{i,j}\right)\sim\text{N}\left(\log (f_j),\frac{\alpha_j^2}{C_{i,j}}\right)\] holds. Let $g_{i,j}$ correspond to the probability density function of $\log (f_{i,j}),$ which lets us write the log-likelihood as follows. Here $A$ is a constant in $f_j$ and thus irrelevant to the optimisation.

\begin{align}
	\ell(f_j) & =\sum^{I-j}_{i=1}\log \left[g_{i,j}\!\left(\log(f_{i,j})\,\middle|\,f_j\right)\right]                                                                                                                                       \\
	          & =\sum^{I-j}_{i=1}\left\{-\frac{1}{2}\cdot\log\left(2\pi\cdot\frac{\alpha_j^2}{C_{i,j}}\right)-\frac{C_{i,j}}{2\cdot\alpha_j^2}\cdot\left[\log\left(f_{i,j}\right)-\log (f_j)\right]^2\right\}                                \\
	          & =A-\frac{1}{2\cdot\alpha_j^2}\sum^{I-j}_{i=1}C_{i,j}\cdot\left[\log(f_{i,j})-\log (f_j)\right]^2
\end{align}

This equals a constant plus the volume-weighted sum of squared log-errors, scaled by the negative constant $-\frac{1}{2\cdot\alpha_j^2}$. It follows that maximising the likelihood is equivalent to minimising the volume-weighted sum of squared log-errors.

\section{Where the accuracy improvement arises}\label{app:stratification}

The pooled means reported in Section~\ref{sec:results} quantify the average \gls{vwga} improvement, but not how the gains are distributed across datasets of differing predictive difficulty. That section showed the performance gap to be greatest for datasets on which both methods predict the \gls{oos} data poorly. To measure the effect, the 362 datasets are grouped into bands according to their \gls{vwa} \gls{rmspe}, separating the triangles the \gls{vwa} fits well from those it fits poorly. Table~\ref{tab:RMSPEStratified} reports each band.

\begin{table}[!htbp]
	\centering
	\renewcommand{\arraystretch}{1.2}
	\setlength{\tabcolsep}{4pt}
	\begin{tabular}{l c c c c c c}
		\toprule
		\rowcolor{gray!15}
		\textbf{\gls{vwa} \gls{rmspe}} & \textbf{Datasets} & \textbf{\gls{vwa}} & \textbf{\gls{vwga}} & \textbf{Reduction} & \textbf{Share} & \textbf{\gls{vwga} better} \\
		\midrule
		\multicolumn{7}{l}{\textbf{Incurred}}                                                                                                  \\
		$[0, 0.1)$         & 149               & 0.0512       & 0.0510        & \phantom{-}0.0002  & \phantom{-}0.3\%  & 83/149 \\
		$[0.1, 0.2)$       & 78                & 0.1454       & 0.1433        & \phantom{-}0.0021  & \phantom{-}1.6\%  & 51/78 \\
		$[0.2, 0.4)$       & 69                & 0.2841       & 0.2684        & \phantom{-}0.0157  & \phantom{-}10.7\% & 50/69 \\
		$[0.4, \infty)$    & 66                & 0.9313       & 0.7976        & \phantom{-}0.1337  & \phantom{-}87.4\% & 61/66 \\
		\midrule
		\multicolumn{7}{l}{\textbf{Paid}}                                                                                                      \\
		$[0, 0.1)$         & 147               & 0.0517       & 0.0521        & -0.0004            & -0.6\%            & 74/147 \\
		$[0.1, 0.2)$       & 85                & 0.1431       & 0.1397        & \phantom{-}0.0035  & \phantom{-}2.8\%  & 51/85 \\
		$[0.2, 0.4)$       & 78                & 0.2760       & 0.2716        & \phantom{-}0.0043  & \phantom{-}3.2\%  & 33/78 \\
		$[0.4, \infty)$    & 52                & 0.8734       & 0.6837        & \phantom{-}0.1898  & \phantom{-}94.6\% & 41/52 \\
		\bottomrule
	\end{tabular}

	\caption{Out-of-sample \gls{rmspe} by \gls{vwa} accuracy band. Reduction is the mean of the per-dataset \gls{vwa} \gls{rmspe} less the \gls{vwga} \gls{rmspe}, so a positive value favours the \gls{vwga}; Share is the band's contribution to the pooled reduction across all 362 datasets.}
	\label{tab:RMSPEStratified}
\end{table}

The improvement is concentrated in the poorly fitted tail. The 52 paid datasets with a \gls{vwa} \gls{rmspe} exceeding 40\% account for 94.6\% of the total reduction in pooled paid error, while the corresponding 66 incurred datasets account for 87.4\% of the pooled incurred reduction. This pattern is consistent with the multiplicative-error argument of Section~\ref{sec:MultErr}, which suggests that the logarithmic transformation should be most beneficial when a small number of unusually large development ratios have a disproportionate influence on the fit.

Two caveats should be noted. First, the bands are defined solely by the \gls{vwa}'s \gls{rmspe}. The resulting sub-samples are therefore selected using the baseline method alone and are not neutral between the competing methods. The numbers above describe where the pooled improvement is concentrated; they are not independent tests of significance. Second, the concentration of the improvement does not imply that the aggregate result is driven by a small number of companies. Removing the 15 datasets with the largest individual reductions strengthens rather than weakens the paired $t$-statistics reported in Section~\ref{sec:results}, because the resulting decrease in the standard deviation of the paired differences exceeds the decrease in their mean.

\section{Dataset selection}\label{app:datasetsSelection}

Four rules governed dataset selection. First, datasets with any negative cumulative values were excluded. Second, any dataset with a full row of zeros was excluded. Third, only triangles with complete histories were used. Fourth, only datasets whose \gls{oos} \gls{mapave} under the \gls{vwa} projection fell below 50\% were retained. The screen was applied to the paid and incurred triangles separately, and a dataset failing on either was excluded. Screening on the \gls{vwa} keeps the criterion independent of the method under evaluation, at the cost of removing the highest-variance datasets, which are those where the \gls{vwga} gains most. Zero cumulative claims were handled by excluding them from calculations, and where every observation in a development period was zero, a development ratio of one was used.

Table~\ref{tab:selectionBreakdown} attributes the exclusions to the rule responsible. A dataset is charged to the first rule it fails, and within a rule the paid triangle is examined before the incurred one; the third rule therefore never charges the incurred triangle, because the two triangles of these datasets always share the same shape.

\begin{table}[!htbp]
	\centering
	\renewcommand{\arraystretch}{1.2}
	\begin{tabular}{l l c c}
		\toprule
		\rowcolor{gray!15}
		         & \textbf{Rule}                        & \textbf{Paid} & \textbf{Incurred} \\
		\midrule
		1        & Negative cumulative value            & 89            & 6                 \\
		2        & Full row of zeros                    & 249           & 2                 \\
		3        & Incomplete history                   & 53            & 0                 \\
		4        & \gls{mapave} at or above 0.5 under \gls{vwa} & 5     & 6                 \\
		\midrule
		         & Total excluded                       & \multicolumn{2}{c}{410}           \\
		         & Retained                             & \multicolumn{2}{c}{362}           \\
		\bottomrule
	\end{tabular}

	\caption{Datasets excluded by each selection rule, of the 772 available.}
	\label{tab:selectionBreakdown}
\end{table}

A full row of zeros indicates an accident year with no paid claims across its entire observed development, which is characteristic of a very small book. The second rule therefore acts largely as a size screen, and the findings of this study should be read as applying to books of material size rather than to the whole population of Schedule P filings. Every retained dataset is a complete ten-year square covering accident years 1998 to 2007.

The datasets used in this study are displayed in Table~\ref{tab:dataUsed}.

\begin{longtable}{@{}l r >{\raggedright\arraybackslash}p{0.74\textwidth}@{}}
	\caption{List of Schedule P datasets used in the study. Each dataset is one company and line of business. The line labels match the columns of Tables~\ref{tab:RMSPE} to~\ref{tab:absMPE}; the corresponding Schedule P file names are \texttt{comauto}, \texttt{medmal}, \texttt{ppauto}, \texttt{wkcomp}, \texttt{prodliab} and \texttt{othliab}.}\label{tab:dataUsed} \\
	\toprule
	\textbf{Line} & \textbf{Datasets} & \textbf{Company codes} \\
	\midrule
	\endfirsthead
	\multicolumn{3}{@{}l}{\footnotesize\itshape Table~\thetable\ continued from the previous page.} \\
	\toprule
	\textbf{Line} & \textbf{Datasets} & \textbf{Company codes} \\
	\midrule
	\endhead
	\midrule
	\multicolumn{3}{r@{}}{\footnotesize\itshape Continued on the next page.} \\
	\endfoot
	\bottomrule
	\endlastfoot
	ComAuto & 92 & 353, 620, 671, 833, 965, 1066, 1090, 1538, 1716, 1767, 2135, 2143, 2208, 2623, 2712, 3240, 4839, 5185, 5320, 5940, 6408, 6459, 6777, 6947, 7080, 8079, 8427, 8672, 10022, 10100, 10308, 10859, 10894, 11118, 11126, 12866, 13439, 13501, 13528, 13587, 13889, 13943, 14044, 14176, 14257, 14370, 14508, 14974, 15024, 15199, 15997, 16748, 17884, 18163, 18309, 18380, 18686, 18767, 19020, 20690, 21172, 21270, 23574, 23663, 25275, 25950, 26077, 26433, 26797, 26905, 27022, 27065, 28436, 28535, 28550, 28886, 29440, 31550, 32301, 32670, 32875, 34606, 35408, 35483, 37036, 38300, 38466, 38733, 40568, 41300, 44415, 44598 \\
	\addlinespace
	MedMal & 8 & 683, 10115, 15865, 31429, 33049, 36277, 36676, 43656 \\
	\addlinespace
	PPAuto & 96 & 43, 353, 460, 620, 671, 965, 1066, 1090, 1538, 1716, 1767, 2003, 2143, 2208, 2259, 3240, 4839, 5185, 5320, 5690, 6947, 7080, 8427, 8672, 10007, 10022, 10204, 10308, 10336, 10783, 11126, 11231, 12360, 13420, 13439, 13501, 13587, 13595, 13641, 13781, 13889, 13943, 14044, 14176, 14257, 14311, 14370, 14443, 14550, 15024, 15172, 15199, 15210, 15407, 15997, 16373, 16799, 17884, 18163, 18309, 18380, 18686, 18791, 19119, 20430, 23574, 25275, 25755, 26077, 26808, 26905, 27022, 27065, 27499, 27766, 29297, 29440, 31062, 31550, 31810, 32387, 33499, 34509, 34592, 34606, 35173, 35408, 37028, 37850, 40550, 40568, 41041, 41459, 42439, 42749, 43494 \\
	\addlinespace
	WorkComp & 61 & 337, 353, 671, 965, 1066, 1538, 1767, 2135, 2712, 3034, 3240, 5010, 5185, 5940, 6408, 6807, 7080, 8672, 10048, 10191, 10385, 10520, 10659, 10699, 10781, 10800, 10859, 11126, 11347, 11703, 12297, 13439, 13501, 13528, 14176, 14257, 14370, 14508, 14575, 14974, 15148, 15199, 16446, 18309, 18380, 18767, 22635, 23140, 23574, 23663, 24017, 26433, 27529, 34576, 37370, 38300, 38687, 38733, 40126, 41300, 41394 \\
	\addlinespace
	ProdLia & 13 & 78, 86, 620, 667, 1538, 2143, 2712, 5185, 6980, 12260, 14257, 23663, 38300 \\
	\addlinespace
	OthLia & 92 & 620, 671, 683, 833, 1066, 1090, 1473, 1538, 1716, 1767, 2003, 2135, 2143, 2208, 2348, 3000, 3085, 3240, 5185, 5320, 5690, 6459, 6777, 6807, 6947, 7625, 8672, 10020, 10657, 11061, 11118, 11126, 11932, 12866, 13439, 13501, 13889, 13919, 14010, 14044, 14176, 14257, 14370, 14753, 14885, 14974, 15024, 15148, 15172, 15210, 15407, 15571, 15768, 15997, 16799, 17043, 17299, 18163, 18380, 18686, 18767, 20690, 23663, 24830, 26077, 26797, 26808, 27022, 27065, 28436, 28550, 30317, 30651, 32670, 33049, 36277, 36315, 38466, 38733, 39861, 40568, 41068, 41300, 41459, 41467, 41580, 42439, 42846, 43826, 43915, 44075, 44598 \\
\end{longtable}

\section*{Funding}
This research was sponsored by Dynamo Analytics.

\bibliographystyle{plainnat}
\bibliography{references}

\end{document}